\documentclass[letterpaper]{article} 
\usepackage{aaai2026}
\usepackage{times}  
\usepackage{helvet}  
\usepackage{courier}  
\usepackage[hyphens]{url}  
\usepackage{graphicx} 
\usepackage{natbib}  
\usepackage{caption} 
\usepackage{algorithm}
\usepackage{algorithmic}

\usepackage{newfloat}
\usepackage{listings}
\DeclareCaptionStyle{ruled}{labelfont=normalfont,labelsep=colon,strut=off} 
\floatstyle{ruled}
\newfloat{listing}{tb}{lst}{}
\floatname{listing}{Listing}
\title{CARES: Collaborative Agentic Reasoning for Error Detection in Surgery}

\author{
    Chang Han Low\textsuperscript{\rm 1},
    Zhu Zhuo\textsuperscript{\rm 1},
    Ziyue Wang\textsuperscript{\rm 1},
    Jialang Xu\textsuperscript{\rm 2},
    Haofeng Liu\textsuperscript{\rm 1},
    Nazir Sirajudeen\textsuperscript{\rm 2},
    Matthew Boal\textsuperscript{\rm 4},
    Philip J. Edwards\textsuperscript{\rm 2},
    Danail Stoyanov\textsuperscript{\rm 2},
    Nader Francis\textsuperscript{\rm 5},
    Jiehui Zhong\textsuperscript{\rm 3},
    Di Gu\textsuperscript{\rm 3},
    Evangelos B. Mazomenos\textsuperscript{\rm 2},
    Yueming Jin\textsuperscript{\rm 1}
}

\affiliations{
    \textsuperscript{\rm 1}National University of Singapore (NUS), Singapore\\
    \textsuperscript{\rm 2}University College London (UCL), UK\\
    \textsuperscript{\rm 3}The First Affiliated Hospital of Guangzhou Medical University, China\\
    \textsuperscript{\rm 4}Gloucestershire Hospitals NHS Foundation Trust, UK\\
    \textsuperscript{\rm 5}The Griffin Institute, UK\\
    ymjin@nus.edu.sg
}

\usepackage{bibentry}

\begin{document}

\maketitle

\begin{abstract}
Robotic-assisted surgery (RAS) introduces complex challenges that current surgical error detection methods struggle to address effectively due to limited training data and methodological constraints. Therefore, we construct MERP (\textbf{M}ulti-class \textbf{E}rror in \textbf{R}obotic \textbf{P}rostatectomy), a comprehensive dataset for error detection in robotic prostatectomy with frame-level annotations featuring six clinically aligned error categories. In addition, we propose CARES (\textbf{C}ollaborative \textbf{A}gentic \textbf{R}easoning for \textbf{E}rror Detection in \textbf{S}urgery), a novel zero-shot clinically-informed and risk-stratified agentic reasoning architecture for multi-class surgical error detection. CARES implements adaptive generation of medically informed, error-specific Chain-of-Thought (CoT) prompts across multiple expertise levels. The framework employs risk-aware routing to assign error task to expertise-matched reasoning pathways based on complexity and clinical impact. Subsequently, each pathway decomposes surgical error analysis into three specialized agents with temporal, spatial, and procedural analysis. Each agent analyzes using dynamically selected prompts tailored to the assigned expertise level and error type, generating detailed and transparent reasoning traces. By incorporating clinically informed reasoning from established surgical assessment guidelines, CARES enables zero-shot surgical error detection without prior training. Evaluation demonstrates superior performance with 54.3 mF1 on RARP and 52.0 mF1 on MERP datasets, outperforming existing zero-shot approaches by up to 14\% while remaining competitive with trained models. Ablation studies demonstrate the effectiveness of our method. The dataset and code will be publicly available.
\end{abstract}


\section{Introduction}
Robotic-Assisted Surgical (RAS) systems enable precise instrument control through small incisions. These systems fundamentally transform surgical procedures by providing enhanced dexterity and superior visualization, resulting in reduced patient trauma, shorter recovery times, and expanded capabilities for complex procedures such as prostatectomies. While RAS enhances surgical precision, it simultaneously introduces challenges and failure modes that differ from traditional surgery, creating new categories of errors from human-machine interaction complexities. Current literature indicates surgical technical errors affect approximately 10\% of RAS procedures worldwide \cite{vincent2001adverse,alemzadeh2016adverse}, contributing to broader surgical adverse event rates where 30\% of patients experience at least one adverse event globally \cite{duclos2024safety,schwendimann2018occurrence}. Half of all adverse events are preventable \cite{vincent2001adverse,healey2002complications}, yet existing error classification guidelines inadequately address RAS systems' unique technological characteristics, creating a critical gap in patient safety infrastructure.

This inadequacy is particularly concerning given the complexity of RAS-specific errors. Surgical errors inevitably manifest when surgeons struggle with suturing techniques, needle placement, tissue alignment, or grasping strength \cite{tang2020objective,gorard2024application}. These errors also encompass a range of manifestations, including tremor, imprecise movements, or coordination difficulties between robotic arms that can cascade into severe complications during procedures requiring precision.

Currently, surgical error detection relies on manual observation and retrospective analysis by experienced surgeons. This is a subjective, time-intensive process demanding comprehensive analysis of timing, spatial coordination, procedural adherence, and technical precision \cite{xu2024sedmamba}. While standardized assessments like Observational Clinical Human Reliability Assessment (OCHRA) offer systematic approaches to evaluate surgical performance by categorizing errors based on their origin and assessing severity \cite{tang2020objective, gorard2024application, eubanks1999objective, curtis2021clinical,pei2025instrument}, practical implementation remains challenging. Extensive time requirements for manual video analysis and annotation underscore the critical need for computer-assisted assessment solutions \cite{boal2024evaluation,maier2017surgical}, highlighting scalability challenges in modern healthcare environments where the demand for surgical quality evaluation is growing.

Existing AI-assisted surgical error detection systems, such as SEDMamba \cite{xu2024sedmamba} and Chain-of-Gesture (COG) \cite{cog} prompting, demonstrate promising capabilities in binary classification. While these systems can effectively detect the presence of surgical errors, they struggle significantly with multi-class classification tasks that distinguish between specific error categories. This challenge is further complicated by the need to train on datasets from different institutions to capture nuanced variations in surgical techniques and error progression.

The complexity of fine-grained surgical error classification requires sophisticated reasoning capabilities. Vision-language models (VLMs) demonstrate strong reasoning and multi-modal understanding across diverse domains. VLMs can process complex visual information with contextual guidance through structured prompting \cite{llavacot,wang2024qwen2}. Recent agentic AI systems further offer promising paradigms for complex reasoning tasks through role-playing and specialized analytical approaches \cite{RP2,RP3}, with multi-agent frameworks showing improvements in medical image analysis by decomposing complex problems into focused perspectives \cite{GOT}. However, surgical procedures' temporal complexity, where errors span multiple time scales, along with inherent domain gaps in surgical knowledge in current VLMs, creates complex analytical challenges that remain unaddressed.

To address these limitations, we first develop, MERP, \textbf{M}ulti-class \textbf{E}rror in \textbf{R}obotic \textbf{P}rostatectomy, a comprehensive robotic prostatectomy dataset with frame-level annotations based on six clinically aligned fine-grained categories streamlined from SEDMamba \cite{xu2024sedmamba}. Building upon this foundation, we propose \textbf{CARES} (\textbf{C}ollaborative \textbf{A}gentic \textbf{R}easoning framework for \textbf{E}rror Detection in \textbf{S}urgery), a novel zero-shot clinically-informed and risk-stratified agentic reasoning architecture for surgical error detection. CARES adaptively generates medically informed, error-specific Chain-of-Thought (CoT) prompts tailored to multiple expertise level, analytical perspectives and error types. Building on these prompts, CARES employs risk-aware routing to assign error tasks to appropriate expertise-matched analytical pathways, where each pathway decomposes error analysis into three specialized agents conducting temporal, spatial, and procedural analysis. Each agent infer using dynamically selected prompts customised to the assigned expertise level and error type, generating detailed reasoning traces which can be reviewed, contrasting with black-box approaches that offer limited insight. CARES enables zero-shot surgical error detection without prior training which alleviates the data scarcity challenge prevalent in surgical AI. The clinical knowledge integration through structured adaptive prompting ensures CARES align with established surgical assessment frameworks, enabling seamless integration with existing surgical training and quality assurance protocols. Our contributions include:
\begin{itemize}
\item We present a novel dataset, MERP, Multi-class Error in Robotic Prostatectomy, with frame-level annotations to facilitate relevant community study.
\item We introduce an adaptive CoT generation pipeline that generates error-specific, medical-informed reasoning protocols across multiple expertise levels and analytical perspectives.
\item We propose the first clinically-informed and risk-stratified agentic reasoning architecture for zero-shot surgical error detection with a risk-aware routing mechanism for dynamic CoT selection and expertise-matched reasoning pathways.
\item We establish a comprehensive benchmark and demonstrate superior performance across two datasets, achieving 54.3 mF1 on RARP and 52.0 mF1 on MERP without fine-tuning, outperforming existing zero-shot approaches by up to 14\% and remaining competitive with trained models.
\end{itemize}

\section{Related Work}
\subsubsection{Surgical Error Detection and Assessment.} Surgical error detection has evolved from manual observational methods to computational approaches, including kinematic analysis using robot telemetry data \cite{gao2014jhu} and video-based systems leveraging visual information \cite{xu2024sedmamba,cog}. OCHRA established systematic error taxonomies for executional and procedural errors that remain influential in current research \cite{tang2020objective,gorard2024application,eubanks1999objective,curtis2021clinical,qin2025structure}. Recent advances focus on video-based approaches leveraging visual information in surgical recordings. The Chain-of-Gesture framework introduced end-to-end error detection using gesture prompting, but relies on fixed templates lacking adaptive reasoning for diverse error types \cite{cog}. State space models like SEDMamba show promise but require large training data and target binary classification, limiting applicability to multi-class error requiring fine-grained analysis \cite{xu2024sedmamba}.

\subsubsection{Medical Vision-Language Models.} Large-scale vision-language models have transformed medical image analysis by enabling zero-shot clinical content understanding. Models such as CLIP \cite{Clip} and BLIP \cite{Blip} demonstrate capabilities in medical image classification and pathology analysis without domain-specific training \cite{Medclip,medclip2,medclip3}. In surgical applications, VLMs show promise for phase recognition, instrument detection and surgical scene understanding \cite{surgclip,surgclip2,surgvlm}. However, surgical error detection presents unique challenges exceeding current VLM capabilities. The temporal nature of surgical errors, contextual understanding of procedural workflows, and nuanced technical assessment requirements surpass single-model approaches. 

\subsubsection{Multi-Agent Systems and Chain-of-Thought Reasoning.} Multi-agent systems have gained attention with the advancement of large language models, where collaborative reasoning often outperforms individual model capabilities \cite{multiagent}. This has driven extensive applications of multi-agent systems in the medical domain \cite{mdagents,medagent,mmedagent,surgraw}. Chain-of-thought prompting enables step-by-step problem decomposition for complex tasks, extended to visual reasoning in vision-language models through static prompt generation \cite{CoT,llavacot, yang2025visionthink,yang_2}. Recent advances explore adaptive prompting and context-aware prompt modification based on task characteristics. However, existing approaches rely on fixed prompting strategies that fail to account for the varying tasks requirements. We introduce one of the first applications of collaborative agent systems to surgical video analysis, combining multi-agent reasoning with dynamically generated chain-of-thought prompting for complexity-adaptive error detection.

\section{Methodology}
\subsection{MERP Dataset Construction}
We present MERP (\textbf{M}ulti-class \textbf{E}rror in \textbf{R}obotic \textbf{P}rostatectomy), a comprehensive multi-class surgical error dataset for robotic prostatectomy with frame-level annotations across six distinct error categories. The dataset comprises 20 surgical videos from 20 patients captured using the da Vinci surgical system. Originally recorded at 25 fps, we subsequently downsampled to 10 Hz for analysis and yielded 111,555 frames. Video durations range from 5 to 10 minutes, encompassing complete suturing processes with varying case complexity. When comparing with task-specific models, the dataset is split at patient level with 16 videos for training and 4 videos for testing.

\setlength{\tabcolsep}{1mm}
\begin{table}[htbp]
\centering
\begin{tabular}{cp{2.7cm}p{4.8cm}}
\hline
\textbf{ID} & \textbf{Error Category} & \textbf{Descriptions} \\
\hline
1 & Multiple Attempts & Repeated needle suturing attempts during procedure. \\
2 & Out of View & Needle or instruments not visible, with danger-dependent severity. \\
3 & Needle Handling & Drops, incorrect grip angles, wrong positioning, poor curve following. \\
4 & Tissue Handling & Tissue damage from poor stabilization and excessive force. \\  
5 & Suture Handling & Thread catching, inadequate throws, entanglement, fraying, snapping. \\
6 & Instrument Control & Poor camera control, inadequate handling, instrument clashing. \\
\hline
\end{tabular}
\caption{Error category descriptions for the MERP dataset.}
\label{tab:error_categories}
\end{table}

\subsubsection{Error Taxonomy and Annotation Framework.}
We employed validated annotation frameworks from OCHRA checklist~\cite{tang2020objective,gorard2024application,eubanks1999objective,curtis2021clinical} and SEDMamba~\cite{xu2024sedmamba}. After extensive consultations with urologists and clinicians, the originally proposed 24 error categories in SEDMamba are systematically consolidated into 6 types shown in Table~\ref{tab:error_categories}. This streamlined taxonomy was developed by analyzing practical utility in surgical training contexts, error frequency patterns, and clinical significance. This process prioritized errors with the highest clinical impact and occurrence frequency, creating a clinically meaningful dataset. Frame-level annotations were conducted using Final Cut Pro, denoting the start and end of every error instance. Two urologists experienced in RAS procedures supervised the annotation process, with consensus established through joint review for all disagreements. Detailed dataset description is available in supplementary.

\subsubsection{Dataset Composition and Error Distribution.}

\begin{table}[htbp]
\centering
\begin{tabular}{p{3.8cm}ccc}
\hline
\textbf{Category} & \textbf{Instances} & \textbf{Frames} & \textbf{\%} \\
\hline
\multicolumn{4}{c}{\textit{\textbf{Binary Error Classification}}} \\
No Error & - & 77,763 & 69.7\% \\
Error & - & 33,792 & 30.3\% \\
\textbf{Total Frames} & \textbf{-} & \textbf{111,555} & \textbf{100\%} \\
\hline
\multicolumn{4}{c}{\textit{\textbf{Multi-class Error Classification}}} \\
\multicolumn{4}{c}{\textit{Breakdown of erroneous frames}} \\
Multiple Attempts & 109 & 5,475 & 16.2\%* \\
Out of View & 102 & 5,422 & 16.0\%* \\
Needle Handling Errors & 351 & 11,472 & 33.9\%* \\
Tissue Handling Errors & 14 & 157 & 0.6\%* \\
Suture Handling Errors & 36 & 3,719 & 11.0\%* \\
Instrument Control Errors & 303 & 7,547 & 22.3\%* \\
\textbf{Total Error Frames} & \textbf{915} & \textbf{33,792} & \textbf{100\%} \\
\hline
\end{tabular}
\caption{MERP Dataset Distribution. * indicates percentages for error categories are relative to total error frames.}
\label{tab:error_distribution}
\end{table}

Table~\ref{tab:error_distribution} presents the MERP dataset distribution across 915 error instances and 33,792 error frames. An instance represents a single error event, which may span multiple consecutive frames. Needle Handling and Instrument Control errors dominate, comprising over 70\% of instances, while Tissue Handling errors are least frequent. The multi-class annotations enable binary classification for error detection tasks. We extended this annotation framework to the existing RARP dataset~\cite{RARP}, converting binary classifications to our 6-category taxonomy.

\begin{figure*}[!t]
\centering
\includegraphics[width=\textwidth]{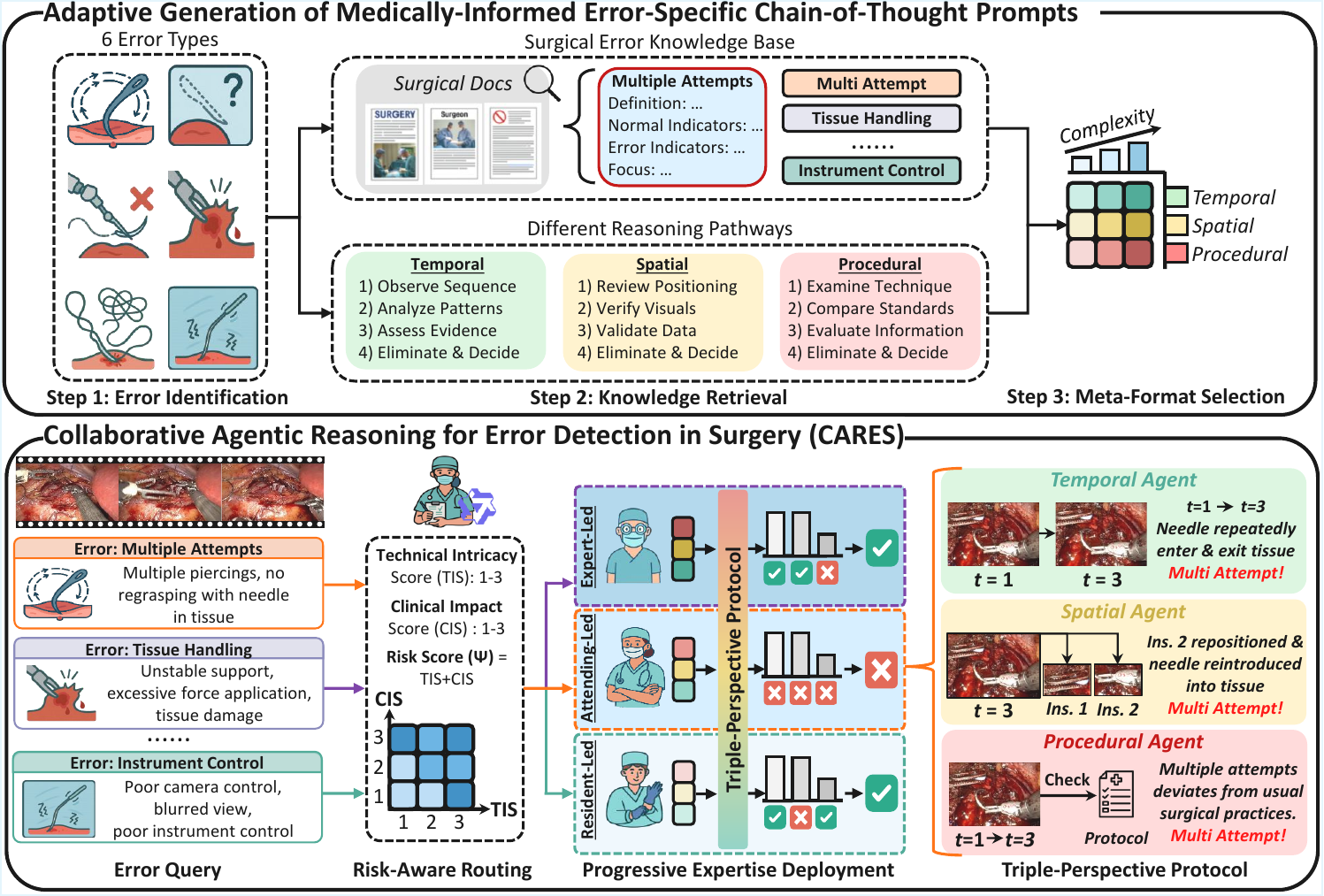}
\caption{Overview of the CARES framework. \textbf{Top}: Adaptive CoT generation integrates error-specific information from surgical error knowledge base with expertise-level reasoning structures and perspective-specific analysis to create specialized reasoning protocols. \textbf{Bottom}: Risk-aware routing computes composite risk scores to deploy expertise-matched reasoning pathways and dynamically select appropriate CoT protocols.}
\label{fig:case}
\end{figure*}

\subsection{Adaptive Generation of Error-Specific CoT Prompts}
\subsubsection{Surgical Error Knowledge Library.}
Surgical error detection requires domain-specific clinical knowledge that extends beyond visual pattern recognition. To enable clinically-informed reasoning and to bridge the medical domain gap in current VLMs, we construct a comprehensive surgical error knowledge library from established clinical resources including OCHRA checklists and expert-validated error taxonomies \cite{xu2024sedmamba,tang2020objective}. Let $\mathcal{E} = \{e_1, e_2, ..., e_6\}$ denote the six error categories defined in our taxonomy: Multiple Attempts, Out of View, Needle Handling Errors, Tissue Handling Errors, Suture Handling Errors, and Instrument Control Errors. For each error type $e_i \in \mathcal{E}$, the system maintains a knowledge repository $\mathcal{K}_i = (D_i, N_i, I_i, F_i)$ where $D_i$ captures clinical definitions and procedural context, $N_i$ documents normal technique indicators and acceptable variations, $I_i$ specifies error indicators and detection criteria, and $F_i$ defines assessment focus areas and critical observation points.

\subsubsection{Expertise-Level Reasoning Structures.}
Surgical error assessment varies across clinical expertise levels \cite{boal2024evaluation,maier2017surgical}. Junior clinicians typically rely on explicit criteria and systematic verification, while experienced practitioners integrate multiple competing hypotheses and subtle contextual factors \cite{expertise1,expertise2,expertise4,expertise5,expertise6}. Role-playing prompts have been shown to activate domain-specific reasoning patterns in language models, motivating our design of three expertise-matched reasoning structure and pathways \cite{RP1,RP2,RP3}. Our framework implements three reasoning pathways reflecting these assessment sophistication levels. Resident-Level Structure ($\mathcal{L}_R$) employs systematic, checklist-based reasoning with conservative interpretation. Attending-Level Structure ($\mathcal{L}_A$) balances structured evaluation with contextual interpretation. Expert-Level Structure ($\mathcal{L}_E$) enables sophisticated pattern synthesis with multi-scale temporal analysis.

\subsubsection{Multi-Perspective Analytical Dimensions with Structured Reasoning.}
Surgical error assessment requires analysis across temporal, spatial, and procedural dimensions. Our framework employs three specialized analytical perspectives, each with tailored CoT reasoning protocols. Temporal Analysis ($\mathcal{A}_T$) examines timing patterns and motion sequences, distinguishing between acceptable timing variations and clinically significant deviations such as multiple attempts or prolonged hesitation indicating procedural difficulties. Spatial Analysis ($\mathcal{A}_S$) evaluates positional accuracy and spatial relationships within the surgical field, distinguishing between technical positioning errors and acceptable anatomical variation. Procedural Analysis ($\mathcal{A}_P$) compares observed techniques against established surgical protocols, accounting for clinically acceptable procedural adaptations. Each perspective generates domain-specific reasoning chains that systematically evaluate evidence while acknowledging the limitations of video-based assessment. This process generates 54 distinct CoT prompts (6 error types $\times$ 3 expertise levels $\times$ 3 analytical perspectives), creating a comprehensive library of reasoning protocols. Each prompt is constructed using the compositional framework:
\begin{equation}
\mathrm{CoT}_{p,x}^{e_i} = \mathcal{L}_x(\mathcal{A}_p(\mathcal{K}_i)), \quad x \in \{R,A,E\}, \quad p \in \{T,S,P\}
\end{equation}

\subsection{Dynamic CoT Selection and Orchestration}
Surgical errors span multiple time scales and require integration of spatial, temporal, and procedural context, creating analytical challenges that vary significantly in both detection difficulty and clinical consequences. Current detection model apply uniform processing regardless of variations, failing to match analytical rigor to risk profiles of different error scenarios. While the above section detailed static but adaptive generation of reasoning protocols, our framework's intelligence lies in dynamically selecting appropriate prompts for each error case. Unlike approaches that apply uniform reasoning regardless of complexity, CARES adapts its analytical approach based on error characteristics and clinical risk. The dynamic risk-aware routing determines which expertise-level pathway to activate during the progressive expertise deployment phase and coordinates the triple-perspective protocol for optimal detection. This ensures complex, high-risk cases receive sophisticated expert-level analysis while routine cases are processed through resident-level protocols, bridging static reasoning pathways with context-aware surgical knowledge base retrieval.

\subsection{Risk-Aware Expertise-Driven Agent Architecture}
\subsubsection{Risk-Aware Routing Mechanism.}
To address the varying analytical complexity and clinical significance of different error scenarios, we design a dual-metric risk assessment. Our system computes a composite risk score for each potential error instance $e_{i}$ in video clip $V$:
\begin{equation}
\Psi(e_{i}) = \mathrm{TIS}(e_{i}) + \mathrm{CIS}(e_{i})
\end{equation}
where $e_{i}$ denotes an instance of error type $i$. The Technical Intricacy Score $\mathrm{TIS} \in \{1, 2, 3\}$ quantifies the detection complexity and the Clinical Impact Score $\mathrm{CIS} \in \{1, 2, 3\}$ evaluates potential patient safety implications. Higher $\Psi(e_{i})$ indicate errors requiring more sophisticated analytical approaches. The routing mechanism assigns each case to expertise-matched analytical pathways based on  $\Psi(e_{i})$:
\begin{equation}
\Theta(e_{i}) = \left\{
\begin{array}{ll}
\mathcal{P}_R & \mathrm{if}~\Psi(e_{i}) \in \{2,3\} \\
\mathcal{P}_A & \mathrm{if}~\Psi(e_{i}) \in \{4,5\} \\
\mathcal{P}_E & \mathrm{if}~\Psi(e_{i}) = 6
\end{array}
\right.
\end{equation}
where $\Theta$ denotes the routing function that assigns error instances to pathways, and $\mathcal{P}_R$, $\mathcal{P}_A$, and $\mathcal{P}_E$ represent Resident-level, Attending-level, and Expert-level pathways, respectively. This allocation ensures complex, high-impact errors receive appropriately sophisticated analysis.

\subsubsection{Expertise-Matched Agent Deployment.}
Each risk-stratified pathway deploys three specialized agents corresponding to temporal, spatial, and procedural analysis through a \textbf{triple perspective protocol}:
\begin{eqnarray}
\mathcal{P}_R &=& \{\mathcal{A}_{T,R}, \mathcal{A}_{S,R}, \mathcal{A}_{P,R}\} \\
\mathcal{P}_A &=& \{\mathcal{A}_{T,A}, \mathcal{A}_{S,A}, \mathcal{A}_{P,A}\} \\
\mathcal{P}_E &=& \{\mathcal{A}_{T,E}, \mathcal{A}_{S,E}, \mathcal{A}_{P,E}\}
\end{eqnarray}

Each agent produces comprehensive reasoning analysis that consolidates to a binary assessment $\mathcal{O}_{p,x}(V) \in \{0,1\}$ for video clip $V$. The triple perspective protocol integrates multi-agent perspectives through:
\begin{equation}
\mathcal{C}_{\mathcal{P}_x}(V) = \alpha_T \cdot \mathcal{O}_{T,x}(V) + \alpha_S \cdot \mathcal{O}_{S,x}(V) + \alpha_P \cdot \mathcal{O}_{P,x}(V)
\end{equation}
where $\mathcal{C}_{\mathcal{P}_x}(V)$ represents the pathway decision score and $\alpha_T > \alpha_S > \alpha_P$ reflects the relative importance of temporal, spatial, and procedural evidence. These weights are empirically calibrated by varying each parameter individually while maintaining others at baseline ($\alpha = 1.0$), isolating individual agent contributions while avoiding exponential search complexity. 
Final decisions use a unified threshold $\theta$ across all pathways:
\begin{equation}
\hat{y}_x = \left\{
\begin{array}{ll}
1 & \mathrm{if}~\mathcal{ C}_{\mathcal{P}_x}(V) > \theta \\
0 & \mathrm{otherwise}
\end{array}
\right.
\end{equation}

The threshold $\theta$ is empirically optimized through systematic evaluation across the range [1.0-3.3] to balance detection sensitivity with false positive control while ensuring robust multi-agent consensus validation.

\section{Experiments}

\subsection{Experimental Setup}

\subsubsection{Implementation Details.} We implement CARES using Qwen2.5-VL as the VLM backbone. All inference are conducted on 1 NVIDIA A6000 GPUs with 48GB VRAM.

\subsubsection{Datasets.} We evaluate framework performance on two datasets with distinct error characteristics, MERP and RARP. For RARP, we applied our 6-category annotation protocol to convert the original binary labels into multi-class classifications. Both datasets use 10-second clips with 1-second stride (9-second overlap), with clips labeled as erroneous if any frame contains error annotations.

\subsubsection{Evaluation Metrics.} Due to inherent class imbalance in surgical data, we report macro-averaged F1 scores (mF1) and balanced accuracy (bACC) as primary metrics. These metrics provide equal weighting across error categories regardless of frequency distribution, ensuring robust assessment of minority class performance. Results represent averages across 5 independent runs, with detailed standard deviations and case studies in supplementary material.

\subsubsection{Baseline Comparisons.} We compare CARES against five state-of-the-art VLMs: InternVL2.5-8B \cite{InternVL25}, InternVL3-9B \cite{Internvl3}, Qwen2.5-VL-7B \cite{Qwen25VL}, LLAVA-OV-7B \cite{LLAVAOV}, and Video-LLAMA-7B \cite{VideoLLama}. All VLM baselines employ zero-shot evaluation protocols with standard prompting and are configured with a maximum token limit of 1024, top-k sampling of 40, and both top-p and temperature at 0.8. For the supervised baselines SEDMamba \cite{xu2024sedmamba} and COG \cite{cog}, we follow the original training settings from their respective papers.

\subsection{Comparative Study}
\setlength{\tabcolsep}{2.5mm} 
\begin{table*}[htbp]
\centering
\begin{tabular}{c|cc|cc|cc|cc|cc|cc}
\hline
\multicolumn{13}{c}{\textbf{RARP Dataset}} \\
\hline
& \multicolumn{2}{c|}{Error 1} & \multicolumn{2}{c|}{Error 2} & \multicolumn{2}{c|}{Error 3} & \multicolumn{2}{c|}{Error 4} & \multicolumn{2}{c|}{Error 5} & \multicolumn{2}{c}{Error 6} \\
Model & mF1 & bACC & mF1 & bACC & mF1 & bACC & mF1 & bACC & mF1 & bACC & mF1 & bACC \\
\hline
COG* & 44.4 & 50.3 & 40.9 & 50.0 & 38.4 & 50.0 & 44.9 & 50.1 & 44.7 & 50.4 & 36.4 & 50.3 \\
SEDMamba* & \underline{51.0} & \underline{51.4} & \textbf{68.6} & \textbf{68.0} & \underline{49.5} & \textbf{53.8} & 44.8 & 50.1 & 46.0 & \underline{50.8} & \underline{49.2} & \underline{53.5} \\
InternVL2.5 & 38.9 & 49.5 & 44.9 & 51.3 & 46.7 & 49.8 & 38.3 & 48.7 & 40.7 & 50.1 & 49.1 & 50.1 \\
InternVL3 & 25.9 & 51.0 & 49.8 & 52.6 & 29.4 & 50.1 & 18.8 & \underline{50.4} & 19.5 & 50.4 & 31.7 & 49.8 \\
Qwen2.5VL & 41.1 & 50.0 & 35.8 & 41.2 & 37.5 & 49.0 & 44.9 & 47.3 & 43.8 & 49.7 & 42.4 & 51.2 \\
LLAVA-OV & 49.4 & 49.6 & 41.7 & 50.0 & 41.9 & 48.3 & \underline{47.9} & 49.0 & \underline{47.8} & 49.1 & 40.5 & 48.1 \\
Video-LLAMA & 48.0 & 50.2 & 43.3 & 49.1 & 47.8 & 48.5 & 45.7 & 45.7 & 47.2 & 47.5 & 48.3 & 49.4 \\
\textbf{CARES} & \textbf{56.3} & \textbf{59.1} & \underline{55.7} & \underline{56.4} & \textbf{49.7} & \underline{51.3} & \textbf{58.3} & \textbf{60.1} & \textbf{49.8} & \textbf{51.6} & \textbf{55.8} & \textbf{58.7} \\
\hline
\multicolumn{13}{c}{\textbf{MERP Dataset}} \\
\hline
& \multicolumn{2}{c|}{Error 1} & \multicolumn{2}{c|}{Error 2} & \multicolumn{2}{c|}{Error 3} & \multicolumn{2}{c|}{Error 4} & \multicolumn{2}{c|}{Error 5} & \multicolumn{2}{c}{Error 6} \\
Model & mF1 & bACC & mF1 & bACC & mF1 & bACC & mF1 & bACC & mF1 & bACC & mF1 & bACC \\
\hline
COG* & \textbf{49.0} & \textbf{50.0} & 49.1 & 50.0 & 46.6 & 50.0 & 49.7 & 50.0 & \underline{49.7} & 50.0 & 47.1 & 50.0 \\
SEDMamba* & \textbf{49.0} & \textbf{50.0} & \textbf{76.2} & \textbf{77.4} & \textbf{61.6} & \textbf{59.0} & \underline{49.9} & 50.0 & \underline{49.7} & 50.0 & 47.5 & 50.2 \\
InternVL2.5 & 14.8 & 46.9 & 49.5 & 50.1 & 43.2 & 50.1 & 43.2 & \textbf{51.2} & 18.8 & 48.6 & 44.5 & 49.5 \\
InternVL3 & 10.8 & 47.2 & 47.7 & 48.1 & 27.5 & 50.5 & 14.0 & 47.3 & 26.0 & \underline{50.3} & 20.2 & 50.1 \\
Qwen2.5VL & 47.4 & \underline{48.9} & 48.8 & 50.0 & 47.7 & 50.0 & 38.4 & \underline{50.7} & 41.4 & 47.6 & \underline{48.8} & \underline{51.6} \\
LLAVA-OV & 44.6 & 46.9 & 48.8 & 50.0 & 42.0 & 48.0 & 45.8 & 47.0 & 49.5 & 50.0 & 42.9 & 50.0 \\
Video-LLAMA & \underline{48.7} & \textbf{50.0} & 43.4 & 48.3 & 48.3 & 50.8 & 40.1 & 48.7 & 40.1 & 49.3 & 46.5 & 47.2 \\
\textbf{CARES} & 47.9 & \underline{48.9} & \underline{54.3} & \underline{54.9} & \underline{55.3} & \underline{56.8} & \textbf{50.2} & 49.7 & \textbf{53.1} & \textbf{54.8} & \textbf{52.1} & \textbf{53.3} \\
\hline
\end{tabular}
\caption{Performance comparison on error detection tasks across RARP and MERP datasets (error IDs per Table 1). Best results in bold, second-best underlined. *Indicates supervised methods requiring model training.}
\label{tab:mainerror_performance}
\end{table*}

\setlength{\tabcolsep}{3mm}
\begin{table}[htbp]
\centering
\begin{tabular}{l|cc|cc}
\hline
& \multicolumn{2}{c|}{\textbf{RARP}} & \multicolumn{2}{c}{\textbf{MERP}} \\
\hline
Model & mF1 & bACC & mF1 & bACC \\
\hline
COG* & 16.9 & 49.9 & 45.9 & 50.0 \\
SEDMamba* & \underline{55.6} & \textbf{62.2} & \textbf{63.3} & \textbf{60.3} \\
InternVL2.5 & 46.8 & 50.2 & 34.5 & 50.0 \\
InternVL3 & 46.0 & 50.1 & 34.3 & 50.1 \\
Qwen2.5VL & 42.8 & 49.3 & 47.8 & 51.2 \\
LLAVA-OV & 37.5 & 47.3 & 32.3 & 49.8 \\
Video-LLAMA & 48.2 & 50.1 & 39.5 & 49.3 \\
\textbf{CARES} & \textbf{57.5} & \underline{59.1} & \underline{53.7} & \underline{55.9} \\
\hline
\end{tabular}
\caption{Performance comparison on binary classification task across RARP and MERP datasets. Best results in bold, second-best underlined. *Indicates supervised methods requiring model training.}
\label{tab:binary_performance}
\end{table}

\subsubsection{Multi-class Error Classification.}
Table~\ref{tab:mainerror_performance} demonstrates the superior performance of CARES in error categories in both datasets. Despite analyzing in the zero-shot setting, CARES substantially outperforms all baseline VLMs and remains competitive against trained models. CARES averages 54.3 mF1 on RARP and 52.0 mF1 on MERP, while baseline VLMs show inconsistent performance with notable drops between datasets. CARES outperforms trained SEDMamba on multiple error types on RARP, while COG performs near chance level. Median standard deviations across 5 runs are consistently lower for CARES (0.6) compared to baseline VLMs (0.8) and supervised methods (1.6).

\subsubsection{Binary Classification Performance.} 
Table~\ref{tab:binary_performance} confirms CARES' effectiveness in binary error detection. CARES achieves competitive performance (RARP: 57.5 mF1, MERP: 53.7 mF1) compared to trained SEDMamba (55.8 mF1 on RARP, 59.8 mF1 on MERP), demonstrating strong capability without training. COG shows poor performance on RARP, while baseline VLMs achieve moderate performance. Median standard deviations across 5 runs are consistently lower for CARES (1.2) compared to baseline VLMs (0.5) and supervised methods (2.3). 

We selected Qwen2.5VL based on superior instruction-following capabilities and low hallucination rates ~\cite{wang2024qwen2}. While other VLMs show hallucination issues or struggle with complex reasoning, Qwen2.5VL demonstrates consistent behavior critical for surgical applications \cite{li2023evaluating,hallu1,hallu2, hallu_new1}.

\subsection{Ablation}

\begin{table}[!ht]
\centering
\begin{tabular}{lcc}
\hline
Method & Avg. mF1 & Avg. bACC \\
\hline
\multicolumn{3}{c}{\textbf{RARP}} \\
\hline
Baseline & 40.9 & 48.1 \\
+ Static CoT & 51.6 (↑10.7) & 53.8 (↑5.7) \\
\quad (Majority Vote) & & \\
+ Dynamic CoT & 51.9 (↑11.0) & 54.0 (↑5.9) \\
\quad (Risk-Aware Routing) & & \\
+ Dynamic CoT & \textbf{54.3 (↑13.4)} & \textbf{56.2 (↑8.1)} \\
\quad (Full CARES) & & \\
\hline
\multicolumn{3}{c}{\textbf{MERP}} \\
\hline
Baseline & 45.4 & 49.8 \\
+ Static CoT & 49.1 (↑3.7) & 51.7 (↑1.9) \\
\quad (Majority Vote) & & \\
+ Dynamic CoT & 50.5 (↑5.1) & 52.4 (↑2.6) \\
\quad (Risk-Aware Routing) & & \\
+ Dynamic CoT & \textbf{52.0 (↑6.6)} & \textbf{53.1 (↑3.3)} \\
\quad (Full CARES) & & \\
\hline
\end{tabular}
\caption{Performance improvement by progressively adding CARES components.}
\label{tab:orchestration_methods}
\end{table}

\subsubsection{Framework Component Contribution Analysis.} Table~\ref{tab:orchestration_methods} presents four incremental configurations isolating each CARES component. Baseline employs single VLM classification. Static CoT introduces multi-agent reasoning with fixed prompts across nine agents (three expertise levels × three perspectives) using majority voting. Dynamic CoT adds risk-aware routing while maintaining majority voting within pathways. Full CARES implements triple-perspective protocol, prioritizing temporal analysis for sequential surgical errors. CoT reasoning provides foundational improvements, with static CoT achieving substantial gains over baseline (+10.7 mF1 on RARP, +3.7 mF1 on MERP). Dynamic prompting with risk-aware routing yields modest but consistent refinements. The triple-perspective protocol delivers the largest incremental improvement (+2.4 mF1 on RARP, +1.5 mF1 on MERP). This progression validates that surgical error detection benefits from structured reasoning sophistication.

\subsubsection{Analysis of Chain-of-Thought Strategies.}

Table~\ref{tab:cot_comparison} evaluates three CoT strategies: baseline classification without reasoning, single-agent generic CoT using "think step by step" prompting \cite{tsbs}, and our medically-informed error-specific CoT with majority voting. Generic CoT provides minimal gains, while medically-informed CoT delivers substantial improvements. This underscores the necessity for domain-tailored prompting strategies.

\setlength{\tabcolsep}{6mm}
\begin{table}[!ht]
\centering
\begin{tabular}{lcc}
\hline
Method & Avg. mF1 & Avg bACC \\
\hline
\multicolumn{3}{c}{\textbf{RARP}} \\
\hline
Baseline & 40.9 & 48.1 \\
Single CoT & 41.5 (↑0.6) & 50.3 (↑2.2) \\
Static CoT & \textbf{51.6 (↑10.7)} & \textbf{53.8 (↑5.7)} \\
\hline
\multicolumn{3}{c}{\textbf{MERP}} \\\hline
Baseline & 45.4 & 49.8 \\
Single CoT & 46.3 (↑0.9) & 48.3 (↓1.5) \\
Static CoT & \textbf{49.1 (↑3.7)} & \textbf{51.7 (↑1.9)} \\
\hline
\end{tabular}
\caption{Performance comparison of Chain-of-Thought prompting strategies. Single CoT uses generic CoT prompt, while Static CoT uses medical domain-informed CoT.}
\label{tab:cot_comparison}
\end{table}

\subsubsection{Hierarchical Role-Playing Expertise Analysis.}  
\begin{figure}[!ht]
    \centering
    \includegraphics[width=0.9\linewidth]{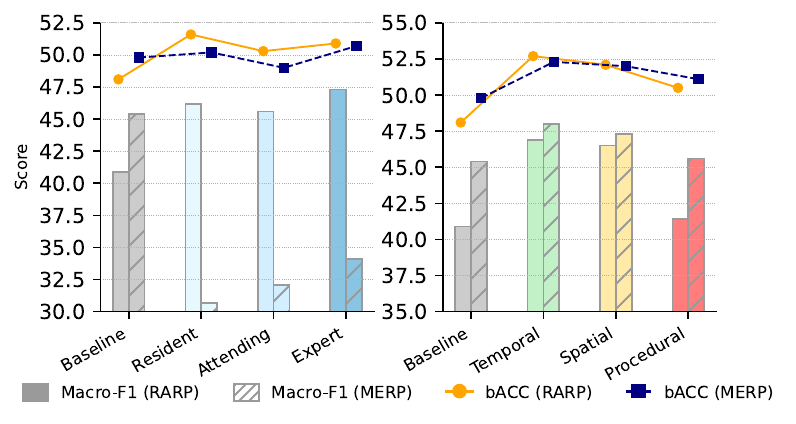}
    \caption{\textbf{Left: Expertise levels.} Performance comparison for Baseline, Resident, Attending, and Expert models. \textbf{Right: Reasoning styles.} Performance comparison for Baseline, Temporal, Spatial, and Procedural variants. 
}
    \label{fig:ablation}
\end{figure}

Figure~\ref{fig:ablation} (Left) evaluates individual expertise levels by incorporating role-playing prompts with medical expertise descriptions without multi-perspective analysis. Expert-level reasoning achieves optimal performance on RARP, but all expertise approaches perform poorly on MERP. This shows that simple role playing lacks robustness for zero-shot deployment.

\subsubsection{Perspective Reasoning and Temporal Focus.} Figure~\ref{fig:ablation} (Right) evaluates individual analytical perspectives using single-focus reasoning without role-playing. Temporal analysis consistently outperforms other perspectives on both datasets (46.9 mF1 RARP, 48.0 mF1 MERP), followed by spatial analysis, with procedural analysis providing complementary insights. These validate our triple-perspective protocol approach, confirming that temporal patterns provide the strongest discriminative signal for detection.

\subsubsection{Alpha Parameter Calibration Analysis.} Figure~\ref{fig:threshold} (Left) reveals distinct optimal operating regions and values. Temporal reasoning achieves peak performance with highest amplification while spatial reasoning benefits from moderate enhancement. This pattern highlights the importance of temporal reasoning for surgical error detection, where dynamic analysis enhances detection compared to static spatial relationships or technique validation. The clear performance peaks demonstrate that reasoning perspectives contribute optimally at different amplification levels.

\subsubsection{Threshold Optimization and Temporal Analysis.} 
\begin{figure}[!ht]
    \centering
    \includegraphics[width=1\linewidth]{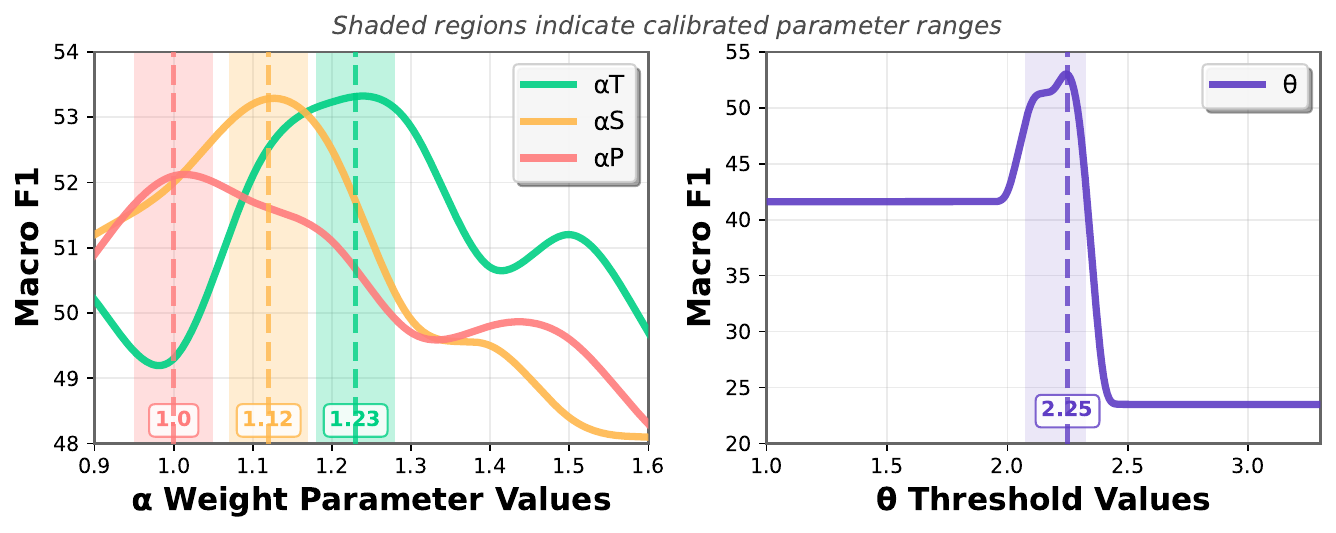}
    \caption{\textbf{Left: Agent Weight Calibration ($\alpha$).} Performance optimization across perspective weights. \textbf{Right: Threshold Optimization ($\theta$).} Performance across decision threshold range for multi-agent consensus.
}
    \label{fig:threshold}
\end{figure}

Figure~\ref{fig:threshold} (Right) shows distinct performance tiers across consensus strategies. Liberal thresholds ($\leq$2.0) suffered from false positives, while basic consensus (2.1-2.2) achieved moderate improvements. Optimal performance at 2.25 enforces temporal agent participation, reflecting surgical errors' dynamic nature. This demonstrates spatial reasoning captures static relationships but misses error progression, while procedural assessment validates standards without evaluating dynamic changes. Error detection requires temporal pattern analysis with spatial and procedural validation.

\section{Conclusion and Future Works}
We introduced CARES, the first zero-shot risk-stratified multi-agent framework for robotic surgical error detection. Our framework demonstrates that medically-informed chain-of-thought reasoning provides foundational improvements, while risk-aware routing and triple-perspective protocol further enhances. Temporal analysis proves most discriminative, confirming error detection requires temporal reasoning over static patterns. CARES achieves competitive performance against trained models despite analyzing in a zero-shot setting, establishing a new paradigm for surgical assessment. Future work will expand to other RAS procedures and other institutions to validate the framework's generalizability across diverse surgical contexts, techniques and error manifestations. Future clinical validations studies with surgeons can be conducted to establish practicality.

\end{document}